\documentstyle[multicol,aps,prl]{revtex} 

\renewcommand{\narrowtext}{\begin{multicols}{2} \global\columnwidth20.5pc}
\renewcommand{\widetext}{\end{multicols} \global\columnwidth42.5pc}
\def\al{\alpha}

\def\ga{\gamma}
\def\de{\delta}

\def\th{\theta}

\def\ka{\kappa}

\def\si{\sigma}

\def\ps{\psi}

\def\De{\Delta}

\def\frac#1#2{{\textstyle{{#1}\over {#2}}}}

\def\ket#1{|{#1}\rangle}

\def\half{{\textstyle{1\over 2}}}
\def\lsim{\mathrel{\rlap{\lower4pt\hbox{\hskip1pt$\sim$}}
    \raise1pt\hbox{$<$}}}
\def\gsim{\mathrel{\rlap{\lower4pt\hbox{\hskip1pt$\sim$}}
    \raise1pt\hbox{$>$}}}
\def\sqr#1#2{{\vcenter{\vbox{\hrule height.#2pt
         \hbox{\vrule width.#2pt height#1pt \kern#1pt
         \vrule width.#2pt}
         \hrule height.#2pt}}}}

\def\hydrogen{H}
\def\antihydrogen{$\overline{\rm{H}}$}

\newcommand{\beq}{\begin{equation}}
\newcommand{\eeq}{\end{equation}}
\newcommand{\bea}{\begin{eqnarray}}
\newcommand{\eea}{\end{eqnarray}}
\newcommand{\rf}[1]{(\ref{#1})}
 
\begin{document}

\title{CPT and Lorentz Tests in Hydrogen and Antihydrogen}
\author{Robert Bluhm,$^a$ V. Alan Kosteleck\'y,$^b$ and
Neil Russell$^b$} 
\address{$^a$Physics Department, Colby College,
          Waterville, ME 04901, U.S.A.}
\address{$^b$Physics Department, Indiana University, 
          Bloomington, IN 47405, U.S.A.}
\date{IUHET 388, COLBY-97-04, April 1998}
\maketitle

\begin{abstract}
Signals for CPT and Lorentz violation at the Planck scale
may arise in hydrogen and antihydrogen spectroscopy.
We show that certain 1S-2S and hyperfine transitions
can exhibit theoretically detectable effects 
unsuppressed by any power of the fine-structure constant.
\end{abstract}

\narrowtext

Experimental and theoretical studies of the spectrum of hydrogen
(\hydrogen) 
have historically been connected 
to several major advances in physics
\cite{H}.
The recent production and observation of antihydrogen (\antihydrogen) 
\cite{oelert,mandel}
makes it plausible to consider a new class 
of spectroscopic measurements 
involving high-precision comparisons 
of the spectra of \hydrogen\ and \antihydrogen\
\cite{ce}.
The two-photon 1S-2S transition 
has received much attention 
because an eventual measurement of the line center 
to about $1$ mHz,
corresponding to a resolution of one part in $10^{18}$,
appears feasible
\cite{hanschICAP}.
It has already been measured 
to $3.4$ parts in $10^{14}$
in a cold atomic beam of \hydrogen\ 
\cite{hansch}
and to about one part in $10^{12}$
in trapped \hydrogen\ 
\cite{cesar}.
Proposed \antihydrogen\ spectroscopic investigations
involve both beam and trapped-atom techniques
\cite{mandel2,gab2}.

We consider here the use of spectroscopy of free 
or magnetically trapped \hydrogen\ and \antihydrogen\ 
to search for CPT and Lorentz violation. 
The discrete symmetry CPT is an invariance 
of all local Lorentz-invariant 
quantum field theories of point particles 
\cite{cpt},
including the standard model 
and quantum electrodynamics (QED).
However,
the situation is less clear for 
a more fundamental theory 
combining the standard model with gravity,
such as string theory,
where spontaneous breaking of these symmetries may occur
\cite{kps}.
Low-energy manifestations 
would be suppressed by a power of the ratio
of the low-energy scale to the Planck scale,
so only a few exceptionally sensitive experiments 
are likely to detect them.

In this work,
we show that effects of this type from the Planck scale
can appear in \hydrogen\ and \antihydrogen\ spectra
at zeroth order in the fine-structure constant.
Moreover,
these effects are theoretically detectable not only  
in 1S-2S lines but also in hyperfine transitions. 

Our analysis is performed in the context of
a standard-model and QED extension 
\cite{ck}
incorporating the idea of spontaneous CPT and Lorentz
breaking at a more fundamental level.
This quantum field theory appears 
at present 
to be the only existing candidate for a consistent extension
of the standard model based on a microscopic theory
of CPT and Lorentz violation.
Desirable features such as energy-momentum conservation,
gauge invariance,
renormalizability,
and microcausality
are expected \cite{ck}.
The theory has been applied to 
photon properties
\cite{ck},
neutral-meson experiments
\cite{kps,ckpv,expt,ak},
Penning-trap tests 
\cite{bkr},
and baryogenesis
\cite{bckp}.

We begin with a study of
the spectra of \it free \rm 
\hydrogen\ and \antihydrogen. 
For this case,
the standard-model extension 
generates a modified Dirac equation
for a four-component electron field $\ps$ 
of mass $m_e$ and charge $q = -|e|$ in 
the proton Coulomb potential $A^\mu = (|e|/4 \pi r, 0)$.
With $i D_\mu \equiv i \partial_\mu - q A_\mu$,
this equation (in units with $\hbar = c = 1$) is
$$\left( i \ga^\mu D_\mu - m_e - a_\mu^e \ga^\mu
- b_\mu^e \ga_5 \ga^\mu 
\right.
\qquad \qquad \qquad \qquad \qquad
$$
\beq
\left. 
- \half H_{\mu \nu}^e \si^{\mu \nu} 
+ i c_{\mu \nu}^e \ga^\mu D^\nu 
+ i d_{\mu \nu}^e \ga_5 \ga^\mu D^\nu \right) \ps = 0
\quad .
\label{dirac}
\eeq
The two terms involving the couplings
$a_\mu^e$ and $b_\mu^e$ violate CPT,
while the three terms involving 
$H_{\mu \nu}^e$, $c_{\mu \nu}^e$, and $d_{\mu \nu}^e$
preserve CPT.
All five couplings break Lorentz invariance
and are assumed to be small
\cite{ck}.
A modified Dirac equation also exists for a free proton
\cite{bkr},
and it contains corresponding couplings
$a_\mu^p$, $b_\mu^p$, $H_{\mu \nu}^p$, $c_{\mu \nu}^p$, 
and $d_{\mu \nu}^p$
\cite{fn0}.

To examine the spectra of free \hydrogen\ and \antihydrogen,
it suffices to perform a perturbative calculation 
in the context of relativistic quantum mechanics.
In this approach,
the unperturbed hamiltonians
and their eigenfunctions are
the same for \hydrogen\ and \antihydrogen.
Moreover,
all perturbative effects
from conventional quantum electrodynamics
are also identical for both systems.
However,
the CPT- and Lorentz-breaking couplings
for the electron and positron 
can provide different hermitian perturbations 
to the hamiltonians describing \hydrogen\ and \antihydrogen.
The explicit forms of these perturbations are found
from Eq.\ \rf{dirac}
by a standard procedure involving charge conjugation
(for \antihydrogen)
and field redefinitions
\cite{bkr}.
Similarly,
CPT- and Lorentz-breaking couplings 
for the proton and antiproton 
generate additional energy perturbations.
These can be obtained to leading order using 
relativistic two-fermion techniques
\cite{D2}.

Let $J=1/2$ and $I=1/2$
denote the (uncoupled) electronic and nuclear angular momenta,
respectively,
with third components $m_J$, $m_I$.
The energy corrections for the basis states $\ket{m_J,m_I}$
can then be calculated perturbatively.
To leading order,
we find the energy corrections 
for spin eigenstates of protons or antiprotons
have the same mathematical form as those 
for electrons or positrons,
except for the replacement of superscripts $e$ with $p$
on the CPT- and Lorentz-violating couplings.

For \hydrogen, 
we find the 1S and 2S levels acquire 
identical leading-order energy shifts. 
They are 
\cite{fn1}
\bea
\De E^{H} (m_J, m_I)
& \approx &
(a_0^e + a_0^p - c_{00}^e m_e - c_{00}^p m_p)
\cr
&&
+ (-b_3^e + d_{30}^e m_e + H_{12}^e) {m_J}/{|m_J|} 
\cr
&&
+ (-b_3^p + d_{30}^p m_p + H_{12}^p) {m_I}/{|m_I|} ~ ,
\label{EHJI}
\eea
where $m_p$ is the proton mass.
For \antihydrogen,
the 1S and 2S levels also acquire 
identical leading-order energy shifts
$\De E^{ \overline{H}}$,
which are given by the expression \rf{EHJI}
with the substitutions
$a_\mu^e \rightarrow - a_\mu^e$,
$d_{\mu \nu}^e \rightarrow - d_{\mu \nu}^e$,
$H_{\mu \nu}^e \rightarrow - H_{\mu \nu}^e$;
$a_\mu^p \rightarrow - a_\mu^p$,
$d_{\mu \nu}^p \rightarrow - d_{\mu \nu}^p$,
$H_{\mu \nu}^p \rightarrow - H_{\mu \nu}^p$.

The hyperfine interaction couples the electron and proton
or positron and antiproton spins.
Denoting the total angular momentum by $F$,
the appropriate basis states become 
linear combinations $\ket{F,m_F}$
of the $\ket{m_J,m_I}$ states.
The selection rules for the two-photon 1S-2S transition
are $\De F = 0$ and $\De m_F = 0$
\cite{cagnac}.
There are thus four allowed 1S-2S transitions 
for both \hydrogen\ and \antihydrogen,
occurring between states with the same spin configuration.
However,
according to Eq.\ \rf{EHJI}
the 1S and 2S states with the same spin configuration
have identical leading-order energy shifts,
so no leading-order effects appear in the frequencies
of any of these transitions.
Thus,
there is no leading-order 1S-2S spectroscopic signal 
for Lorentz or CPT violation in 
either free \hydrogen\ or free \antihydrogen
\cite{fn2}.

The dominant subleading energy-level shifts
involving the CPT- and Lorentz-breaking couplings 
in free \hydrogen\ and \antihydrogen\
arise as relativistic corrections of order $\al^2$.
These differ for some of the 1S and 2S levels
and therefore could in principle lead
to observable effects.
For example,
the term proportional to $b_3^e$ in Eq.\ \rf{dirac} 
produces a frequency shift in the 
$m_F = 1 \rightarrow m_{F^\prime} = 1$ line
relative to the $m_F = 0 \rightarrow m_{F^\prime} = 0$ line
(which remains unshifted),
given by 
$\de \nu^H_{1S-2S} \approx - \al^2 b_3^e / 8 \pi$.
Similarly,
the proton-antiproton corrections are also 
suppressed by factors at least of order
$\al^2 \simeq 5\times 10^{-5}$.
The suppression factors reduce the signals
in both free \hydrogen\ and free \antihydrogen\
to levels that could in principle be excluded
by results from feasible $g-2$ experiments.
In fact,
the estimated attainable bound 
\cite{bkr}
on $b_3^e$ from 
electron-positron $g-2$ experiments performed 
with existing technology
would suffice to place a bound 
of $\de \nu^H_{1S-2S} \lsim 5$ $\mu$Hz
on observable shifts of the 1S-2S frequency 
in free \hydrogen\ from the electron-positron sector. 
This is below the resolution of the 1S-2S line center.
Although no Penning-trap $g-2$ experiments on protons 
and antiprotons have yet been performed,
bounds attainable in such experiments would
also yield tighter constraints on the proton-antiproton
parameters than would be obtained in 1S-2S measurements.

At first sight,
it may seem surprising that bounds from
$g-2$ experiments can constrain observable effects 
in comparisons of 1S-2S
transitions in free \hydrogen\ and \antihydrogen.
The conventional figure of merit for CPT breaking 
in $g-2$ experiments 
involving the difference of the electron
and positron $g$ factors is 
$r_g = |g_{e^-} - g_{e^+}|/g_{\rm av} \lsim 2 \times 10^{-12}$
\cite{pdg},
which is six orders of magnitude weaker
than the idealized resolution of the 1S-2S line,
$\De \nu_{1S-2S}/\nu_{1S-2S} \simeq 10^{-18}$.
However,
the use of $r_g$ in Penning-trap 
$g-2$ experiments can be inappropriate
in the present theoretical context
\cite{bkr}.
The relevant physical issues are the
absolute frequency resolution and the sensitivity 
to CPT- and Lorentz-violating effects.
The absolute frequency resolution 
in $g-2$ measurements is approximately 1 Hz,
which is about three orders of magnitude poorer 
than the idealized 1S-2S line-center resolution.
In constrast,
the $g-2$ experiments involve spin-flip transitions
that induce direct sensitivity to $b_3^e$,
whereas the 1S-2S transitions in free \hydrogen\ or \antihydrogen\ 
are sensitive only to the suppressed combination $\al^2 b_3^e/8\pi$.
The resulting bound on $b_3^e$ from 1S-2S comparisons 
is thus about two orders of magnitude weaker 
than that from electron-positron $g-2$ experiments.
The above discussion suggests 
that experiments with \hydrogen\ and \antihydrogen\ 
might obtain tighter bounds by studying transitions 
between states with different spin configurations.
Accomplishing this requires the presence of external fields.

We next consider spectroscopy of \hydrogen\ or \antihydrogen\ 
confined within a magnetic trap
with an axial bias magnetic field,
such as an Ioffe-Pritchard trap \cite{ip}.
This situation is directly relevant to 
proposed experiments
\cite{gab2}.
Denote each of the four 1S and 2S hyperfine Zeeman levels 
in order of increasing energy 
in a magnetic field $B$ with the labels 
$\ket{a}_n$, $\ket{b}_n$, $\ket{c}_n$, $\ket{d}_n$,
with $n=1$ or $2$, 
for both \hydrogen\ and \antihydrogen.
For \hydrogen,
the mixed-spin states are given in terms of the 
basis states $\ket{m_J,m_I}$ as
\bea
\ket{c}_n &=& \sin \th_n \ket{-\half,\half} +
\cos \th_n \ket{\half,-\half}
\quad ,
\nonumber\\
\ket{a}_n &=& \cos \th_n \ket{-\half,\half} -
\sin \th_n \ket{\half,-\half}
\quad .
\label{a}
\eea
The mixing angles $\th_n$ 
depend on the principal quantum number $n$
and obey
$\tan 2 \th_n \approx (51 {\rm ~mT})/n^3B$.
Prior to excitation,
the states that remain confined in the trap
are the low-field seekers,
$\ket{c}_1$ and $\ket{d}_1$. 
However, 
spin-exchange collisions 
$\ket{c}_1 + \ket{c}_1 \rightarrow \ket{b}_1 + \ket{d}_1$ 
lead to a loss of population 
of the $\ket{c}_1$ states over time,
resulting in confinement of primarily $\ket{d}_1$ states.

Transitions between the unmixed-spin states 
$\ket{d}_1$ and $\ket{d}_2$ are field independent 
for small values of the magnetic field.
It would therefore seem natural to compare 
the frequency $\nu^H_d$ 
for the 1S-2S transition $\ket{d}_1 \rightarrow \ket{d}_2$ 
in \hydrogen\ 
with the frequency $\nu^{\overline{H}}_d$ 
for the corresponding transition in \antihydrogen. 
However,
since in \hydrogen\ the spin configurations of the 
$\ket{d}_1$ and $\ket{d}_2$ states are the same,
there are again no unsuppressed frequency shifts.
The same result holds for \antihydrogen.
Thus,
to leading order we find
$\de \nu^H_d = \de \nu^{\overline{H}}_d \simeq 0$.

A theoretically interesting alternative 
would be to consider instead the 1S-2S transition
$\ket{c}_1 \rightarrow \ket{c}_2$ in \hydrogen\ 
and the corresponding transition in \antihydrogen. 
The idea would be to take advantage of the mixed 
nature of these states in a nonzero magnetic field.
The $n$ dependence in the hyperfine splitting
produces a spin-mixing difference between the 1S and 2S levels,
giving an unsuppressed frequency shift 
in 1S-2S transitions 
between the $\ket{c}_1$ and $\ket{c}_2$ states:
\beq
\de \nu_c^H \approx
-\ka (b_3^e - b_3^p - d_{30}^e m_e 
+ d_{30}^p m_p - H_{12}^e + H_{12}^p)/2\pi ~,
\label{nucH}
\eeq
where $\ka\equiv \cos 2\th_2 - \cos 2\th_1$.
The analogous 1S-2S frequency shift 
$\de \nu_c^{\overline{H}}$ for \antihydrogen\ 
in the same magnetic field can also be found.
The hyperfine states in \antihydrogen\ 
have opposite positron and antiproton spin assignments 
relative to those of the electron and proton in \hydrogen,
so $\de \nu_c^{\overline{H}}$ is given by an expression 
identical to that for $\de \nu_c^H$ in Eq.\ \rf{nucH}
except that the signs of $b_3^e$ and $b_3^p$ are changed.
The frequencies $\nu_c^H$ and $\nu_c^{\overline H}$
depend on spatial components of Lorentz-violating couplings
and so would exhibit diurnal variations in the comoving 
Earth frame.
There would also be a nonzero instantaneous difference 
$\De \nu_{1S-2S,c} \equiv \nu_c^H 
- \nu_c^{\overline{H}} \approx - \ka (b_3^e - b_3^p)/\pi$
for measurements made in the same magnetic trapping fields.
The value of this difference would depend 
on the 1S-2S spin-mixing difference controlled by $\ka$
\cite{fn3}.

The theoretical gain in sensitivity to CPT and Lorentz violation 
of the $\ket{c}_1 \rightarrow \ket{c}_2$ transition
relative to that of $\ket{d}_1 \rightarrow \ket{d}_2$
would be of order $4/\al^2 \simeq 10^5$.
However,
since the 1S-2S transition
$\ket{c}_1 \rightarrow \ket{c}_2$
in \hydrogen\ and \antihydrogen\ is field dependent,
any experiment would need to overcome Zeeman broadening 
due to the inhomogeneous trapping fields.
For example,
at $B\simeq 10$ mT the 1S-2S linewidth 
for the $\ket{c}_1 \rightarrow \ket{c}_2$
transition is broadened to over 1 MHz 
for both \hydrogen\ and \antihydrogen\
even at a temperature of $100 \mu$K.
Existing techniques might partially mitigate this effect,
but the development of other experimental methods would appear 
necessary to attain resolutions on the order of the natural linewidth.

As an alternative to optical spectroscopy of the 1S-2S line,
we consider frequency measurements of transitions in the 
hyperfine Zeeman effect.
Since transitions between $F = 0$ and $F^\prime = 1$
hyperfine states have been measured with accuracies 
better than $1$ mHz in a hydrogen maser
\cite{ramsey},
hyperfine transitions in masers and 
in trapped \hydrogen\ and \antihydrogen\
are interesting candidates
for tests of CPT and Lorentz symmetry.

In the 1S ground state of hydrogen,
all four hyperfine levels acquire energy shifts
due to CPT- and Lorentz-violating effects.
Each energy shift contains an identical contribution 
$a_0^e + a_0^p -c_{00}^e m_e -c_{00}^p m_p$
that leaves transition frequencies unaffected.
The remaining spin-dependent contributions to the
energy shifts are 
\bea
\De E_a^H &\simeq&  
\hat\ka (b_3^e - b_3^p - d_{30}^e m_e 
+ d_{30}^p m_p - H_{12}^e + H_{12}^p)
\quad ,
\nonumber\\
\De E_b^H &\simeq& 
b_3^e + b_3^p - d_{30}^e m_e 
- d_{30}^p m_p - H_{12}^e - H_{12}^p
\quad ,
\nonumber\\
\De E_c^H &\simeq& -\De E_a^H
\quad , \qquad
\De E_d^H \simeq - \De E_b^H
\quad ,
\label{abcd}
\eea
where $\hat\ka \equiv \cos2 \th_1$.
In zero magnetic field $\hat\ka =0$, 
so the energies of $\ket{a}_1$ and $\ket{c}_1$ are unshifted.
However, 
$\ket{b}_1$ and $\ket{d}_1$ acquire equal and opposite 
energy shifts.
The degeneracy of the three $F=1$ ground-state
hyperfine levels is therefore lifted even for $B=0$
\cite{fn4}.
For example,
the transitions 
$\ket{d}_1 \rightarrow \ket{a}_1$
and $\ket{b}_1 \rightarrow \ket{a}_1$ 
exhibit an unsuppressed and diurnally varying 
frequency difference
$|\De \nu_{d-b}^H| \approx
|b_3^e + b_3^p - d_{30}^e m_e - d_{30}^p m_p
- H_{12}^e - H_{12}^p|/\pi$.
With nonzero values of the magnetic field,
all four hyperfine Zeeman levels acquire energy shifts.
For $\ket{a}_1$ and $\ket{c}_1$,
they are controlled by the spin-mixing parameter $\hat\ka$,
increasing with $B$ and attaining
$\hat\ka \simeq 1$ when $B \simeq 0.3$ T.

The conventional \hydrogen\ maser operates on the
field-independent $\si$ transition 
$\ket{c}_1 \rightarrow \ket{a}_1$
in the presence of a small ($B \lsim 10^{-6}$ T) magnetic field.
Since $\hat\ka \lsim 10^{-4}$ in this case,
the leading-order effects due to
CPT and Lorentz violation in high-precision
measurements of the maser line 
$\ket{c}_1 \rightarrow \ket{a}_1$ are suppressed. 
However,
the frequency difference between the 
field-dependent transitions $\ket{d}_1 \rightarrow \ket{a}_1$
and $\ket{b}_1 \rightarrow \ket{a}_1$
is shifted relative to its usual value 
by $\De \nu_{d-b}^H$,
and the associated diurnal variations 
would provide an unsuppressed signal 
of CPT and Lorentz violation.
Although measurements of this difference 
with existing techniques are possible in principle,
the frequency resolution would be significantly
less than that of the field-independent $\si$ line 
because of broadening due to field inhomogeneities.
Moreover,
an unambiguous resolution of this signal would 
require distinguishing it from possible backgrounds 
arising from residual Zeeman splittings.

The issue of background splittings could in principle
be addressed by direct comparison of transitions
between hyperfine Zeeman levels in \hydrogen\ and \antihydrogen.
Furthermore,
the frequency dependence on the magnetic field
could be eliminated to first order 
by using a field-independent transition point.
One possibility might be to perform 
high-resolution radiofrequency spectroscopy
on the $\ket{d}_1 \rightarrow \ket{c}_1$ transition
in trapped \hydrogen\ and \antihydrogen\
at the field-independent transition point $B \simeq 0.65$ T.
Among the experimental issues to consider would be 
Doppler broadening and 
that the relatively high bias field implies 
potentially larger field inhomogeneities,
so cooling to temperatures of 100 $\mu$K
with a good signal-to-noise ratio 
and a stiff box shape for the trapping potential 
may be needed to obtain frequency resolutions of order 1 mHz.

At this bias-field strength,
the electron and proton spins in
the state $\ket{c}_1$ are highly polarized with
$m_J = 1/2$ and $m_I = - 1/2$.
The transition $\ket{d}_1 \rightarrow \ket{c}_1$ is
effectively a proton spin-flip transition.
We find the frequency shifts 
for \hydrogen\ and \antihydrogen\ are 
$\de \nu_{c \rightarrow d}^H \approx
(-b_3^p + d_{30}^p m_p + H_{12}^p)/\pi$
and 
$\de \nu_{c \rightarrow d}^{\overline{H}} \approx 
(b_3^p + d_{30}^p m_p + H_{12}^p)/\pi$.
The frequencies
$\nu_{c \rightarrow d}^H$ 
and
$\nu_{c \rightarrow d}^{\overline{H}}$ 
would exhibit diurnal variations.
Their instantaneous difference,
\beq
\De \nu_{c \rightarrow d} \equiv 
\nu_{c \rightarrow d}^H - \nu_{c \rightarrow d}^{\overline{H}}
\approx - 2 b_3^p / \pi
\quad ,
\label{nudiff}
\eeq
could provide a direct, clean, and accurate test 
of CPT-violating couplings $b_3^p$ for the proton
\cite{fn5}.

Relevant figures of merit for the various 
direct and diurnal-variation signals 
described in this work can be introduced in analogy
with those for Penning-trap tests
\cite{bkr}.
As one example,
a possible figure of merit for the signal in Eq.\ \rf{nudiff}
would be
\bea
r^H_{rf,c \rightarrow d} & \equiv &
{|({\cal E}_{1,d}^H - {\cal E}_{1,c}^H)
- ({\cal E}_{1,d}^{\overline{H}} - {\cal E}_{1,c}^{\overline{H}})|}/
{{\cal E}_{1,{\rm av}}^H}
\nonumber \\
&\approx &
2\pi |\De \nu_{c \rightarrow d}| /m_H
\quad ,
\label{rrf}
\eea
where ${\cal E}_{1,d}^H$, ${\cal E}_{1,c}^H$ and the
corresponding quantities for \antihydrogen\ each denote
a relativistic energy in a ground-state hyperfine level.
The mass $m_H$ is the atomic mass of \hydrogen.
Thus, 
for example,
attaining a frequency resolution of about 1 mHz 
corresponds to an estimated upper bound of 
$r^H_{rf,c \rightarrow d} \lsim 5 \times 10^{-27}$.
The limit on the CPT- and Lorentz-violating coupling $b_3^p$ 
would then be $|b_3^p| \lsim 10^{-18}$ eV,
which is about three orders of magnitude better
than estimated attainable bounds 
\cite{bkr}
from $g-2$ experiments in Penning traps
and over four orders of magnitude better
than bounds attainable from 1S-2S transitions
\cite{fn6}.

In summary,
we have shown that spin-mixed 1S-2S 
and spin-flip hyperfine spectroscopic signals
for Lorentz and CPT violation 
appear in \hydrogen\ or \antihydrogen\ atoms 
confined in a magnetic trap.
These signals are unsuppressed by any power
of the fine-structure constant 
and would represent observable consequences of
qualitatively new physics originating at the Planck scale.

This work is supported in part by the U.S.\ D.O.E.\
under grant number DE-FG02-91ER40661 and by the N.S.F.\ 
under grant number PHY-9503756.

\end{multicols}

\begin{references}
\bibitem{fn0}
In experiments with \hydrogen\ and \antihydrogen,
suitable field redefinitions can eliminate
various combinations of these couplings,
which thus must be unobservable
\cite{ck}.
To show this explicitly,
we keep all couplings 
in what follows.
\bibitem{fn1}
Note that comparisons of results from different experiments
may necessitate inclusion of geometrical factors associated with 
differing spin-quantization axes.
\bibitem{fn2}
This result is consistent with the demonstration 
in Ref.\ \cite{bkr}
that observable CPT-violating effects must also involve
CT violation and spin-flip processes.
\bibitem{fn3}
Theoretically optimal situations would have maximal $\ka$,
which occurs at $B \simeq 0.01$ T when $\ka \simeq 0.67$.
\bibitem{fn4}
The removal of the $\ket{b}$-$\ket{d}$ degeneracy 
at zero field is compatible with Kramer's theorem
since the Lorentz-violating coefficients 
in Eq.\ \rf{abcd} also violate T.
This splitting might be experimentally detected 
by seeking quantum beats at the difference frequency.
\bibitem{fn5}
All experiments described here are sensitive 
only to spatial components of CPT-violating couplings.
For sensitivity to purely timelike components such as 
$b_0^e$,
a boost would be required.
This can enhance CPT- and Lorentz-violating effects
\cite{ak}.
Although the proposed experiments 
\cite{mandel2}
to measure the fine structure and Lamb shift
with a relativistic beam of \antihydrogen\
are expected to have poorer resolutions
than the others discussed above,
bounds on $b_0^e$ and $b_0^p$ may be attainable.
\bibitem{fn6}
In comparison,
note that the frequency resolution 
of the most accurate clock-comparison experiments,
which also constrain certain types of Lorentz violation,
lies below 1 $\mu$Hz
\cite{hd}.
However,
the theoretical interpretation of these experiments 
is more difficult because relatively complex nuclei are involved,
and in any event no leading-order bounds are obtained on $b_3^p$
in isolation from other couplings 
\cite{kla}.
\end{references}
\end{document}